\documentclass[a4paper,11pt]{article}
\usepackage{pos}
\usepackage{lineno}

\usepackage{upgreek}
\usepackage[capitalise]{cleveref}
\usepackage{hyperref}
\usepackage{multicol}
\usepackage{wrapfig}
\usepackage{booktabs}
\usepackage{enumitem}
\usepackage{xspace}

\usepackage[]{siunitx}

\DeclareSIUnit[quantity-product = ]\percent{\char`\%}
\DeclareSIUnit{\depth}{\gram\per\centi\meter\squared}

\usepackage{tikz}
\usepackage{calc}
\usetikzlibrary{matrix,calc,shapes,decorations.pathreplacing,arrows.meta,decorations.text,hobby,through,decorations.markings,decorations.pathmorphing,patterns,math,tikzmark,positioning,fit}

\usepackage{pgfplots}

\def\Offline{\mbox{$\overline{\text 
      {Off}}$\hspace{.05em}\protect\raisebox{.4ex} 
    {$\protect\underline{\text{line}}$}}\xspace}

\newcommand{\backgroundgrid}[2]{ 
  \begin{scope}[shift={(0.05,0.05)}]
    #1
    \fill[fill=white] (-1,0.4) rectangle (0,1.4);
    \draw[step=0.2,black,thin,#2] (-1,0.4) grid (0,1.4);
  \end{scope}
}

\newcommand{\hexagonn}[3]{ 
  \begin{scope}[shift={(#2,#3)}]
    \draw (30:#1) -- (90:#1) -- (150:#1) -- (210:#1) -- (270:#1) -- (330:#1) -- cycle;
  \end{scope}
}

\newcommand{\hexgrid}[1]{ 
    \foreach \x in {-3, -2, ..., 3} {
        \foreach \y in {-3, -2, ..., 3} {
            \pgfmathsetmacro{\t}{{(\UX * \x + \VX * \y)^2 + (\VY * \y)^2}}
            \pgfmathparse{\t < #1 ? int(1) : int(0)}

            \ifnum\pgfmathresult=1 
            \hexagonn{\HEXSIZE}{{\UX * \x + \VX * \y}}{{\UY * \x + \VY * \y}}
            \fi
        }
    }
}

\tikzset{
    clip even odd rule/.code={\pgfseteorule}, 
    invclip/.style={
        clip,insert path=
            [clip even odd rule]{
                [reset cm](-\maxdimen,-\maxdimen)rectangle(\maxdimen,\maxdimen)
            }
    }
}
\pgfdeclarelayer{ft}
\pgfdeclarelayer{bg}
\pgfsetlayers{bg,main,ft}

\usepackage{layouts}

\newcommand{\xmax}{\ensuremath{X_{\text{max}}}}

\newcommand{\numAddFeatures}{2}
\newcommand{\numAddFeaturesTxt}{two}

\title{ 
Reconstruction of the depth of the shower maximum of air showers with the SD-750 surface detector of the Pierre Auger Observatory using neural networks
}

\ShortTitle{Reconstruction of \xmax{} with the SD-750 using neural networks}

\author*[a]{Steffen Hahn}
\onbehalf{for the Pierre Auger Collaboration$^b$}

\affiliation[a]{ 
Karlsruhe Institute of Technology - Institute for Astroparticle Physics, \\
Hermann-von-Helmholtz-Platz 1, 76344 Eggenstein-Leopoldshafen, Germany}

\affiliation[b]{ 
Observatorio Pierre Auger,\\
Av. San Martín Norte 304, 5613 Malargüe, Argentina\\
Full author list: {\rm\url{https://www.auger.org/archive/authors_icrc_2025.html}}}

\emailAdd{spokespersons@auger.org}

\abstract{ 
The origin of ultra-high-energy cosmic rays (UHECRs) is one of the intriguing mysteries in astroparticle physics.
In order to identify their sources, we need precise knowledge of the mass composition of UHECRs.
The direct detection of UHECRs is not feasible at energies above \qty{0.1}{\peta\eV}, necessitating the use of mass-sensitive observables of extended air showers induced by UHECRs interacting with the atmosphere.
One way to achieve high statistics for these mass-sensitive observables is to use ground-based detector arrays, such as the Surface Detector (SD) of the Pierre Auger Observatory.
The SD consists of three sub-arrays of independent detector stations arranged in triangular grids with different spacings.
Recently, it has been shown that neural networks (NNs) can extract mass-sensitive observables from data taken by the SD-1500, the largest sub-detector of the SD.
In this contribution, we demonstrate the feasibility of using NNs to reconstruct a high-level shower observable, the depth of the shower maximum, from data simulated for and observed by the SD-750, the second-largest detector array nested within the SD-1500.
A simulation study shows that the SD-750 NN exhibits behavior similar to that of an SD-1500 NN and outperforms the latter in the energy range $[1, 10)\, \unit{\exa\eV}$.
Moreover, we show that, after performing a correction and calibration procedure, the predictions of the SD-750 NN are consistent with the measurement of the depth of the shower maximum obtained by the Fluorescence Detector of the Pierre Auger Observatory.
}

\ConferenceLogo{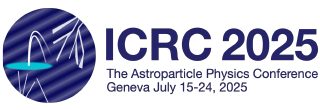}

\FullConference{39th International Cosmic Ray Conference (ICRC2025)\\
 15–24 July 2025\\
Geneva, Switzerland\\}

\begin{document}
\maketitle

\section{Introduction}
\label{sec:introduction}
The Pierre Auger Observatory is a hybrid detector designed to observe extensive air showers (EAS) induced by ultra-high-energy cosmic rays (UHECRs) with several sub-detector systems.
One of the most important sub-detectors is the Surface Detector (SD) of the Pierre Auger Observatory. 
The SD consists of {$\sim$}1660 autonomous detector stations, each of which is equipped with multiple detector systems.
The core of these stations is the water-Cherenkov detector (WCD).
The WCDs stochastically record the weighted densities of the secondary particles from particle cascades that reach the ground, known as the shower footprint.
The detector stations are arranged in three triangular grids with different inter-detector distances.
The grids are named SD-$D$, where $D$ is the distance between the nearest neighbors in meters, referred to without units.
Nested within the main SD array, the SD-1500, the SD-750 is the second largest grid of detector stations.
One of the main objectives of the Pierre Auger Observatory is to measure the mass composition of UHECRs.
Knowledge of the mass composition simultaneously provides insights into the physics at macroscopic scales, such as the sources of UHECRs, and at microscopic scales, such as the high-energy proton-air cross section. 
Due to their scarcity~\cite{AugerSpectrum}, the direct detection of UHECRs around \qty{1}{\exa\eV} is not feasible.
Analyses in this regime depend on so-called `mass-sensitive observables' (MSOs), such as the depth of the shower maximum \xmax{}~\cite{HMModel}, which can be determined indirectly from the measurement of the shower footprint.

Since exploiting the complex spatio-temporal information of the shower footprint is non-trivial, as shown in~\cite{UniversalitySignals}, we mitigate the need for modeling to signal distributions on the ground by using neural networks (NNs). 
An NN-based approach allows us to directly relate the spatio-temporal information of the shower footprint to any MSO. 
This contribution is a feasibility study focused on extracting \xmax{} from the shower footprints of vertical air showers that were simulated for and observed by the WCDs of the SD-750.
It is a low-energy extension of the approach discussed in~\cite{AugerNN1,AugerNN2} using the employed NN architecture (AixNet) as a blueprint.
Using the SD-750 greatly increases the amount of available statistics in the energy range of $[0.1, 3) \, \unit{\exa\eV}$, enabling searches for breaks in the \xmax{} spectra.
\subsection{Neural network architecture}
\label{ssec:neuralNetworkArchitecture}
Mass-sensitive observables, such as the depth of the shower maximum \xmax{}, are properties of the shower.
Therefore, we aim for an event-by-event inference using information from an SD-750 measurement.
Each event is represented by two 3D tensors $I_i$ of the shapes $(F, F, C_i)$, where $C_i$ is the number of channels, e.g., time bins of signal measurement, and $F$ is the size of the encoded shower footprint.
Essentially, the triangular grid of the detectors triggered by an EAS is mapped to the $F{\cdot}F$ positions of the tensors.
The encoding procedure and the preprocessing follow the recipes described in~\cite{Hahn,showerFootprintEncoding}.
The first input tensor $I_0$ contains up to \num{120} bins ($C_0 = 120$) of the signals $S(t)$ detected by the triggered WCDs.
The second input tensor $I_1$ contains \numAddFeaturesTxt{} additional observables ($C_1 = \numAddFeatures$).
We set $F = 7$ as the spatial size of the tensors.

\begin{figure}
  \centering
  \begin{tikzpicture}[x=1.5cm,y=1.5cm,remember picture]
    \tikzstyle{var}=[minimum size=15pt,inner sep=0pt]
    
    \begin{scope}[shift={(0,-0.1)}]
    \begin{scope}
    \backgroundgrid{\backgroundgrid{\backgroundgrid{\backgroundgrid{}{solid}}{dotted}}{dotted}}{dotted}
    \fill[fill=white] (-1,0.4) rectangle (0,1.4);
    \draw[step=0.2,black,thin,fill=white,solid] (-1,0.4) grid (0,1.4);
    \fill[fill=white,opacity=0.7] (-0.5,0.9) circle (0.4);
    \node (signals) at (-0.5,0.9) {$\hat S \left(t\right)$};
    \draw[Latex-Latex] (0.1, 0.3) -- (0.35, 0.55) node [midway, below, sloped] {\small 120};
    \end{scope}
    
    \node[] (compress) at (1, 0.3) {info''};
    \node[draw,circle] (conv1) at (1, 0.9) {sNN1};
    \node[] (compress) at (1, 1.5) {``compress};
    \path[-Latex] (0.25,0.9) edge (conv1);
    \path[-Latex] (conv1) edge (1.75,0.9);
    
    \begin{scope}[shift={(2.75,0)}]
    \backgroundgrid{\backgroundgrid{\backgroundgrid{\backgroundgrid{}{solid}}{dotted}}{dotted}}{dotted}
    \fill[fill=white] (-1,0.4) rectangle (0,1.4);
    \draw[step=0.2,black,thin,fill=white,solid] (-1,0.4) grid (0,1.4);
    \fill[fill=white,opacity=0.7] (-0.5,0.9) circle (0.4);
    \node (signals) at (-0.5,0.9) {$\tilde S (b)$};
    \draw[Latex-Latex] (0.1, 0.3) -- (0.35, 0.55) node [midway, below, sloped] {\small 12};
    \end{scope}
    
    \end{scope}

    \coordinate (aboveplus) at (3.5, 0.9);
    \node[draw,circle] (plus) at (3.5, 0) {$\oplus$};

    \begin{scope}[shift={(0,+0.1)}]
    \begin{scope}[shift={(0,-1.8)}]
    \backgroundgrid{}{solid}
    \fill[fill=white] (-1,0.4) rectangle (0,1.4);
    \draw[step=0.2,black,thin,fill=white] (-1,0.4) grid (0,1.4);
    \fill[fill=white,opacity=0.7] (-0.5,0.9) circle (0.4);
    \node (adds) at (-0.5,0.9) {$\hat A$};
    \end{scope}
    
    \coordinate (belowplus) at (3.5, -0.9);
    
    \node[] (descA1) at (1.85, -0.6) {2 add. observables};
    \node[] (descA2) at (1.85, -1.2) {station state \& rel. trigger time};
  


    \draw (0.2,-0.9) -- (belowplus);
    \end{scope}
    
    \draw (3.2,0.9) -- (aboveplus);
    \path[-Latex] (3.2,0.8) -- (aboveplus) edge (plus);
    \path[-Latex] (0.2,-0.9) -- (belowplus) edge (plus);

    \begin{scope}[shift={(5.0,-0.9)}]
    \backgroundgrid{\backgroundgrid{\backgroundgrid{\backgroundgrid{\backgroundgrid{\backgroundgrid{}{solid}}{solid}}{solid}}{dotted}}{dotted}}{dotted}
    \fill[fill=white] (-1,0.4) rectangle (0,1.4);
    \draw[step=0.2,black,thin,fill=white,solid] (-1,0.4) grid (0,1.4);
    \draw[Latex-Latex] (0.1, 0.3) -- (0.45, 0.65) node [midway, below, sloped] {\small $12 + 2$};
    
    
    \fill[fill=white,opacity=0.7] (-0.5,0.9) circle (0.4);
    \node (signals) at (-0.5,0.9) {$\hat F$};
    \end{scope}
    
    \path[-Latex] (plus) edge (4.0,0);
    
    \node[] (sptext1) at (6.3, 0.6) {``spatial};
    \node[draw,circle] (spatial) at (6.3, 0) {sNN2};
    \node[] (sptext2) at (6.3, -0.6) {analysis''};
    
    \node[] (sptext1) at (5.0, -1.5) {};
    
    \coordinate[] (end) at (7.3, 0);
    \node[] at (7.8, 0) {$X_\text{max}$};

    \path[-Latex] (5.35,0) edge (spatial);
    \path[|-Latex] ($(spatial) + (0.7,+0.0)$) edge (end);
    
    
    \draw[-|] (spatial) edge ++(0.6, 0.0);

\end{tikzpicture}
  \caption{\label{fig:KAneArchitecture} 
    Sketch of the NN architecture: 
    The NN is comprised of two sub-NNs: sNN1 and sNN2.
    Essentially, sNN1 extracts \num{12} features from the \num{120} signal bins recorded by the WCDs.
    These features are then combined with \numAddFeaturesTxt{} additional features (see \cref{ssec:simulationDataSets}) via tensor concatenation.
    Sub-NN sNN2 correlates the spatial information stored in these features to predict \xmax{}.
  }
\end{figure}
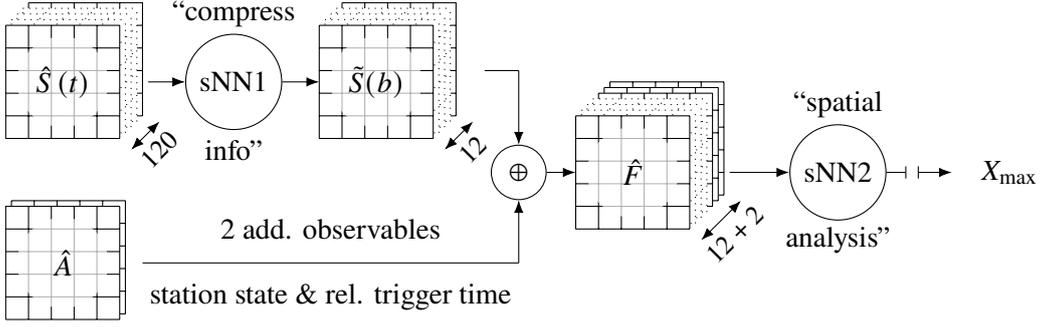

The basic architecture of the NN, sketched in \cref{fig:KAneArchitecture}, consists of two sub-NNs, denoted as sNN1 and sNN2.
The sub-NN sNN1 acts as a feature extraction algorithm compressing the \num{120} signal bins into a \num{12}-dimensional feature vector using long short-term memory layers~\cite{lstm}.
Thus, the output of sNN1 is a tensor of shape $(7,7,12)$.
All input signals are processed in the same way.
Namely, the same weights are used for each spatial position in the encoding.
Positions without a working station are not propagated through sNN1. 
The output of sNN1 is combined with $I_1$ via tensor concatenation yielding a new tensor of shape $(7,7,14)$ used as the input of sNN2.
The sub-NN sNN2 correlates the spatial information of the recorded shower footprint.
It consists of six layers of 2D convolutions which conserve the spatial size of the tensors.
The input of each convolution is added to the output creating so-called shortcuts.
The output is then passed through two blocks, each consisting of a residual layer followed by a 2D convolution that reduces the spatial dimensions.
Finally, the output is flattened and used as the input for the next NN layer: a two-layer perceptron that returns a single prediction for \xmax{}.
All components are implemented using PyTorch~\cite{pytorch}.


\subsection{Simulated data sets and input data}
\label{ssec:simulationDataSets}
Our simulation data set consists of detector simulations based on air shower simulations detected by an ideally working and perfectly regular detector array arranged like the SD-750.
The air shower library consists of an equal mix of air showers induced by proton, helium, oxygen, iron, and tellurium primaries.
Air showers induced by tellurium are simulated using CORSIKA 77560 and air showers induced by the remaining primaries are simulated using CORSIKA 77420~\cite{CorsikaRef}.
In both cases the hadronic interactions are simulated using EPOS LHC~\cite{eposlhc}.
Please note that the tellurium primaries are primarily used to increase the statistics for smaller \xmax{} values.
The detector response is simulated using \Offline~\cite{OfflineRef}.
The data set comprises of Monte Carlo (MC) shower energies, $E_{\text{MC}}$, within $E_{\text{MC}} \in [0.1, 10) \, \unit{\exa\eV}$, which follow an $E^{-1}$ distribution.
The MC zenith angles, $\theta_{\text{MC}}$, are within $\theta_{\text{MC}} \in [\ang{0}, \ang{60})$ and are uniformly distributed in $\sin^2 \theta_{\text{MC}}$.
The energy and zenith ranges contain events that lie outside of the full efficiency of the SD-750 \cite{SD750eff}.
Each simulated air shower is used to generate \num{6} detector responses by randomizing the impact point of the shower core.
The air showers are simulated (in equal parts) for four different atmospheric conditions related to those expected during January, March, August, and September at the observatory.
In total, the simulation data set contains ${\sim}\num[exponent-product=\ensuremath{{\cdot}}]{5e5}$ detector simulations. 

We randomly split our full data set into three sets: a training set and two test sets, using an $8{:}1{:}1$ split.
The first test data set is used to select the best NN from an ensemble of NNs trained under similar conditions.
The second test data set is used for the quality control plot in~\cref{sec:performanceOnSimulations}.
In this way, we reduce the risk of over-interpreting fluctuations in the first data set.
During NN training, we use \qty{10}{\percent} of the training data set for validation.

For each event, \num{768} signal measurements are stored for each triggered WCD and are referred to as signal bins.
The first input tensor $I_0$ contains up to \num{120} signal bins starting from the reconstructed start bin, $b_\text{s}$, of the signal and ending at $\min(b_\text{s} + 120, b_\text{e})$ where $b_{\text{e}}$ is the reconstructed end bin. 
Note that the signal bins of multiple triggered WCDs are not synchronized with each other.
The exact trigger time corresponding to the reconstructed start bin is saved separately. 
Since the signal values span three orders of magnitude, the signal $S(t)$ is logarithmically rescaled as described in~\cite{AugerNN1}.
The second input tensor, $I_1$, contains \numAddFeaturesTxt{} additional features that provide supplementary information about the EAS measurement.
The two features are a map containing the status of the WCDs and the trigger times relative to the station reporting the highest signal~\cite{AugerNN1}. 
The status values for each position in the map are \num{-1} for not working, \num{0} for not triggered, \num{1} for triggered, \num{2} for high-gain saturation, and \num{3} for low-gain saturation.

\subsection{Training procedure}
\label{ssec:trainingProcedure}
Due to the challenging composition measurement of UHECRs, we favor a low bias for the different primary types over high resolution.
During training, we reinforce this by separating the loss calculation for EAS induced by different primaries and by adding an additional bias penalty term to each of these sub-losses.
The loss for a mini-batch $\text{B}$ is
\begin{equation}\label{eq:KAneLoss}
  \mathcal L = \frac{1}{\sum_{{\text{Pr}} \in \text{B}}} \sum_{\text{Pr} \in \text{B}} \left[ \frac{1}{\sum_{i \in \text{Pr}}} \sum_{i \in \text{Pr}} (y_i - p_i)^2 + \left( \frac{1}{\sum_{i \in \text{Pr}}} \sum_{i \in \text{Pr}} [y_i - p_i] \right)^2 \right],
\end{equation}
where $i$ is the sample index, $\text{Pr}$ is a primary (e.g., iron) found in the batch $\text{B}$, $y_i$ is the label, and $p_i$ is the NN prediction.
The first term in \cref{eq:KAneLoss} is a standard mean squared error loss function.
The second term penalizes the total bias of the showers induced by one of the primaries, steering the NN training towards a minimum that exhibits a lower bias for the different primary types.

Occasionally, SD detector stations exhibit malfunctions during operation.
In order to obtain a more accurate footprint distribution, these glitches are mimicked during training by randomly removing stations from the shower footprint and setting their station status to \num{-1}.
During training, we used a turn-off rate of \qty{4}{\percent}, ensuring that at least three stations remain active for each event. 
The SD-750 is relatively small compared to the SD-1500.
Especially at higher energies, shower footprints are not fully contained within the array.
Therefore, the status (see \cref{ssec:simulationDataSets}) of any position in the encoding that is outside of the SD-750 is also set to \num{-1}.

The training process uses Adam~\cite{adamopt} for weight updates.
Training starts with a learning rate of \num[exponent-product=\ensuremath{{\cdot}}]{1.2e-3} using \num{128} events per batch for up to \num{200} epochs.
During training, the learning rate is reduced by a factor of \num{.99} after each epoch.
If the loss on the validation data set (see~\cref{ssec:simulationDataSets}) does not improve for five epochs, the learning rate is reduced by a factor of \num{0.8}.
The training is stopped early if the loss on the validation data set does not improve for eight epochs or if the learning rate drops below \num{e-5}.

\section{Performance on simulations}
\label{sec:performanceOnSimulations}
\begin{figure}
  \centering
  \includegraphics{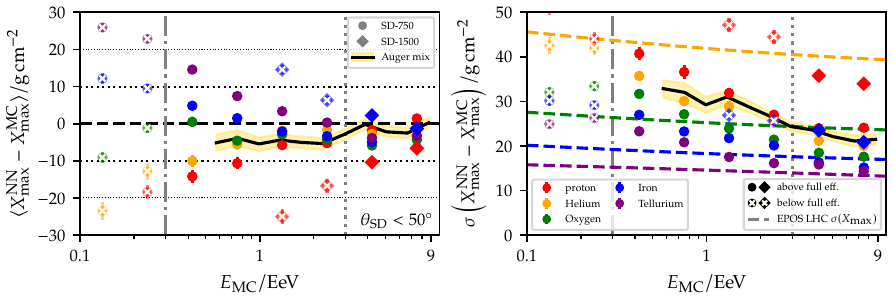}
  \caption{\label{fig:bias_res_vs_energy} 
    First (\emph{left}) and second (\emph{right}) moments of the difference between the predicted and true \xmax{} as a function of the Monte Carlo energy $E_{\text{MC}}$ for showers of the second test data set (see \cref{ssec:simulationDataSets}) simulated in the January and March atmospheres.
    The colors indicate the primary type.
    The colored, dashed lines in the \emph{right} panel depict the width of the true \xmax{} distribution for EPOS LHC~\cite{eposlhc}.
    The black line depicts the bias when using the AugerMix composition for EPOS LHC~\cite{augerMix}.
    The diamonds show the first and second moments for an SD-1500 NN sharing the same architecture as the SD-750 NN.
  }
\end{figure}

To gauge the performance of the NN in simulations, we examine the first and second moments of the difference between the predicted and true \xmax{}, separated for the different primary particle type as a function of the energy (see~\cref{fig:bias_res_vs_energy}) for the January and March atmospheres. 
We interpret the first moment as the bias and the second moment as the resolution of the NN.
Both moments show similar trends to those of AixNet for the SD-1500~\cite{AugerNN1}.
Namely, the bias for different primaries decreases and the resolution improves with increasing energy.
For energies around \qty{0.3}{\exa\eV} the predictions for proton-induced EAS are too low, and for iron-induced EAS are too high exhibiting an averaging-towards-the-mean behavior.
The bias for the proton- and iron-induced EAS is comparable to the average difference of the expected \xmax{} values for proton and iron primaries in EPOS LHC.
Above \qty{1}{\exa\eV}, the bias decreases below \qty{10}{\depth} for all primaries.
At around \qty{3}{\exa\eV}, the biases of the heavy and light primaries cross.
The negative global bias is due to the atmospheres used.
A similar plot for the August and September atmospheres shows a positive global bias at high energies.
The absolute difference between both atmospheres is around \qty{10}{\depth}.
Above \qty{0.3}{\exa\eV}, The resolution improves almost monotonically with increasing energy reaching \qty{\sim 20}{\depth} for the highest energy.
Using the AugerMix composition for EPOS LHC~\cite{augerMix}, we find that the NN predictions exhibit only a weak dependence on the energy which is below \qty{5}{\depth} per decade in energy.
Comparing the moments to that of an NN trained for the SD-1500 using a similar architecture, shows a reduction of the bias and an improvement in resolution in the energy range of $[1, 10) \, \unit{\exa\eV}$.


\section{Performance on measurements}
\label{sec:performanceOnMeasurements}
Since the NN predictions depend strongly on the WCD signals, we reduce the impact of saturation effects and biases resulting from imperfect geometry reconstruction by selecting a high-quality subset from the SD-750 data set.
Inspired by the selection used in~\cite{AugerNN2}, we only consider events with reconstructed core positions for which the distance to the central station lies within the interval $[175, 500] \, \unit{\m}$.
In addition, to minimize edge effects from being too close to the border of the phase space used during training, we only consider events with $\theta < \ang{50}$.
Moreover, we restrict the data set to events with a reconstructed energy $E_{\text{SD}}$ greater than \qty{0.5}{\exa\eV}, which corresponds to the lowest energy bin in the FD \xmax{} spectrum \cite{thomasXmax}.

Due to imperfections in the shower and detector simulations, a direct application of the NN predictions (see~\cref{ssec:trainingProcedure}) introduces biases, as seen in~\cite{AugerNN1}. 
We compensate for the differences between the simulations and the measurements by following the procedure described in~\cite{AugerNN1,AugerNN2}.
First, we correct for effects that were not adequately accounted for in the detector simulation. 
Then, we calibrate the corrected NN predictions by comparing them with the direct \xmax{} measurement on a subset of events detected simultaneously by the FD and the SD-750.

In the simulations, all of the WCDs behave identically.
However, due to aging effects and PMT fluctuations, the detector response of the PMTs in the WCDs differs.
To counteract this difference, we apply the correction described in~\cite{footprintStand} to the recorded signals.
This correction changes the shape of the time signals, making them more simulation-like.
In addition, we correct for any difference between the simulation and the measurement that introduces nonphysical dependencies in the predictions, e.g., on the atmospheric temperature, by using the following procedure:
(1) identify a nonphysical dependence on variable $x_i$, e.g., atmospheric pressure, by looking at the average predictions as a function of $x_i$, (2) try to find a simple model $f_i(x_i; C_{i}, \vec a_{i})$  with the parameters $\vec a_{i}$ and a constant offset $C_{i}$ that captures the relationship found, and (3) choose an anchor point $x_i^\text{a}$ that represents either the expected value in simulations or the most likely value estimated from the distribution of $\{x_i\}$.
Assuming all corrections are independent, we merge them into a global fit model
absorbing all $C_{i}$ into a global offset, $C_{\text{cor}}$.
We restrict ourselves to using only linear or sinusoidal models for each $f_i$.
After performing a least squares fit to all events in the high-quality data set, we set $C_{\text{cor}} = 0$ and add the model values to the NN \xmax{} predictions. 
Compared to the SD-1500 study~\cite{AugerNN2}, there is a stronger dependence on the reconstructed zenith angle.
For $E_{\text{SD}} > \qty{3}{\exa\eV}$, the bias ranges from \qty{-5}{\depth} to \qty{5}{\depth} across low and high zenith angles.

\begin{figure}
  \centering
  \includegraphics{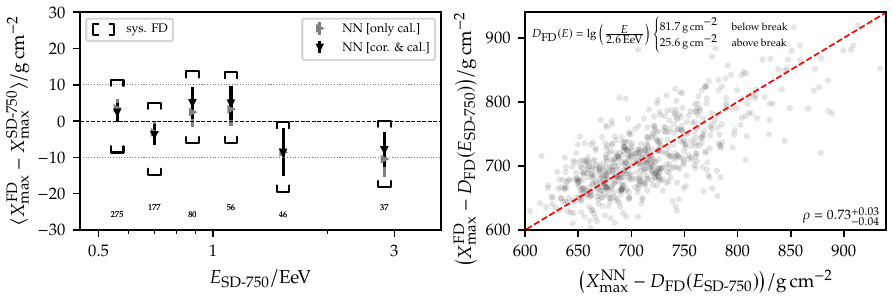}
  \caption{\label{fig:calibrationSD750} 
    Average of event-by-event difference (\emph{left}) and correlation (\emph{right}) of \xmax{} values measured by FD and predicted by the NN for showers observed simultaneously by the FD and the SD-750.
    The average difference in the \emph{left} panel is shown for the NN predictions that are only calibrated (\emph{gray}) and for those that are corrected and then calibrated (\emph{black}).
    The systematic uncertainties reported by FD are added as a reference.
    In the correlation plot on the \emph{right}, we subtract the FD elongation rate, $D_\text{FD}$, to remove the energy dependence, e.g., breaks of \xmax{}~\cite{thomasXmax}.
    The red, dashed line in the \emph{right} panel is the identity line.
    For $E_\text{SD} > \qty{0.5}{\exa\eV}$, the linear correlation $\rho$ between the NN predictions and the FD measurements is \num{0.73}.
    The \qty{95}{\percent} uncertainty on $\rho$ is estimated via bootstrapping.
  }
\end{figure}

The ground signal distributions of the simulations and the measurements differ (e.g.,~\cite{muonPuzzle}).
For AixNet, this introduced a constant bias when comparing high-quality FD measurements with NN predictions.
A similar bias for the predictions was observed in the SD-750 NN predictions when a similar comparison was performed for events that were simultaneously observed by the FD and SD-750.
Assuming that the bias is constant, we find the calibration constant $C_{\text{cal}} = \qty[separate-uncertainty=true]{24.5 \pm 1.6}{\depth}$. 
After the correction and the calibration the difference between the measurement and the prediction is below \qty{\pm 10}{\depth} (see \cref{fig:calibrationSD750}).
The correlation between the FD measurement and NN predictions is \num{0.73} when the elongation rate reported by FD~\cite{thomasXmax} is removed from both \xmax{} values.


\begin{figure}
  \centering
  \includegraphics{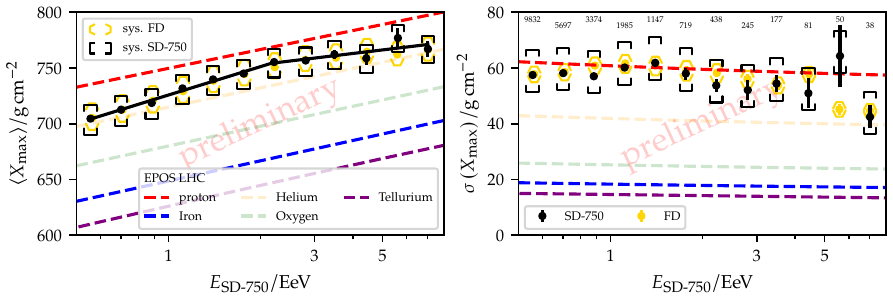}
  \caption{\label{fig:comparisonToFD} 
    Mean (\emph{left}) and standard deviation (\emph{right}) of the NN predictions (\emph{black}) and the FD measurement (\emph{gold}, \cite{thomasXmax}) of \xmax{} as a function of the reconstructed energy.
    The vertical lines through the points show the statistical uncertainties.
    The half-open, convex lines show the systematic uncertainties.
    The dashed lines indicate the expected value of the first and second moments for showers induced by primaries when using EPOS LHC.
    The solid line in the \emph{left} panel is the best fit for the first moment using a piecewise linear function with a single breakpoint.
  }
\end{figure}

Since the calibration procedure is performed with the FD, we assume that all systematic uncertainties of the FD \xmax{} measurement are inherited by the NN prediction.
We add the uncertainty on $C_{\text{cal}}$ and \qty{5}{\depth} for the mass composition bias in~\cref{fig:bias_res_vs_energy} in quadrature to this and use the result as the systematic uncertainty for the mean \xmax{} values.
Because no detector simulations were available for other hadronic interaction models, the systematic uncertainties of the standard deviation are approximated using the uncertainties reported in~\cite{AugerNN1}.
The first and second moments of the corrected and calibrated NN predictions align with the FD measurement within uncertainties (see \cref{fig:comparisonToFD}).
There is only one significant deviation for the second moment in the penultimate energy bin.
When we perform a least squares fit to the means using a piecewise linear function with a single breakpoint, we obtain the elongation rates of \qty[separate-uncertainty=true]{84.8 \pm 10.4}{\depth} and \qty[separate-uncertainty=true]{33.0 \pm 10.3}{\depth} for each decade below and above the break, which occurs at \qty[separate-uncertainty=true]{2.2 \pm 0.5}{\exa\eV}.


\begin{figure}
  \centering
  \includegraphics{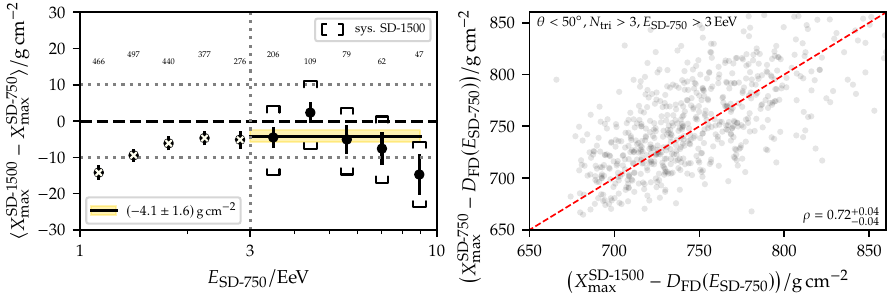}
  \caption{\label{fig:comparisonToSD1500} 
    Average difference between the uncorrected and uncalibrated \xmax{} predictions of the SD-1500 and SD-750 NNs as a function of the SD-750 energy (\emph{left}) and correlation of the predictions (\emph{right}), for events detected by both detectors.
    Only events that triggered more than three stations in the SD-1500 are considered.
    The red, dashed line in the \emph{right} panel is the identity line.
    In the interval $[2, 9) \, \unit{\exa\eV}$, the difference in the NN predictions remains nearly constant.
    The linear correlation, $\rho$, between both NN predictions reaches \num{0.75}.
    The \qty{95}{\percent} uncertainty on $\rho$ is estimated via bootstrapping.
  }
\end{figure}

Since the SD-750 is nested within the SD-1500 there is a subset of EASs which is detected by both arrays simultaneously.
In~\cref{fig:comparisonToSD1500}, we compare the NN predictions of the SD-750 NN and SD-1500 NN, which was used in \cref{fig:comparisonToSD1500}, for the SD-750 energy estimate in the energy range $[1, 10) \, \unit{\exa\eV}$ and for events that triggered more than three SD-1500 stations.
The SD-1500 NN is the one used for the comparison in \cref{fig:comparisonToSD1500}.
In the interval $[2, 9) \, \unit{\exa\eV}$ the difference between the \xmax{} predictions remain relatively constant.
Below \qty{2}{\exa\eV} and above \qty{9}{\exa\eV}, the bias decreases.
Above \qty{3}{\exa\eV}, the total bias between both predictions amounts to \qty[separate-uncertainty=true]{-4.1 \pm 1.6}{\depth}, which is consistent with zero when the systematic uncertainties are taken into account.
Nevertheless, after correcting for the FD elongation, both predictions are linearly correlated by \num{0.72}.


\section{Conclusion}
\label{sec:conclusion}

Although the shower footprints recorded by the SD-750 are more restricted due to its smaller size, the NN presented in this work shows a behavior similar to that of AixNet on the SD-1500 \cite{AugerNN1,AugerNN2}.
In simulations, the bias and resolution improve with increasing energy, surpassing the performance of a similar SD-1500 NN in the energy range $[1, 10) \, \unit{\exa\eV}$ (see \cref{fig:bias_res_vs_energy}).
After a correction and calibration procedure inspired by \cite{AugerNN1}, the first and the second moments of the \xmax{} predictions of the NN are comparable to the FD measurement (see~\cref{fig:comparisonToFD}).
Moreover, comparing the predictions of the SD-750 and SD-1500, NNs indicates that the approach is self-consistent and that the methodology used in~\cite{AugerNN1,AugerNN2} may be extendable to lower energies (see~\cref{fig:comparisonToSD1500}).

\newcommand{\etalpub}{\emph{et al.}}
\newcommand{\augercollabpub}{\etalpub{}{\tiny{${}^{\dagger}$}}}

\renewcommand{\bibsection}{\section*{\refname \hspace{0.5cm} {\normalsize\normalfont\scriptsize $^{\dagger}$ Pierre Auger Collaboration}}}

\begin{multicols}{2}
{\scriptsize\setlength{\columnsep}{3pt}\setlength{\parskip}{0pt}\setlength{\bibsep}{2pt}

}
\end{multicols}

\clearpage

\par\noindent
\textbf{The Pierre Auger Collaboration}\\

\begin{wrapfigure}[8]{l}{0.12\linewidth}
\vspace{-2.9ex}
\includegraphics[width=0.98\linewidth]{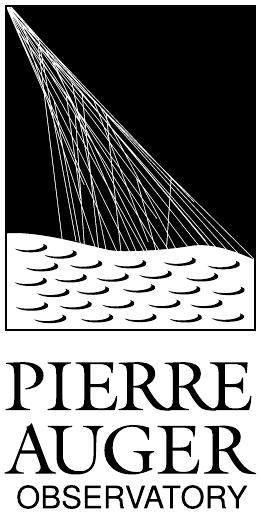}
\end{wrapfigure}
\begin{sloppypar}\noindent
A.~Abdul Halim$^{13}$,
P.~Abreu$^{70}$,
M.~Aglietta$^{53,51}$,
I.~Allekotte$^{1}$,
K.~Almeida Cheminant$^{78,77}$,
A.~Almela$^{7,12}$,
R.~Aloisio$^{44,45}$,
J.~Alvarez-Mu\~niz$^{76}$,
A.~Ambrosone$^{44}$,
J.~Ammerman Yebra$^{76}$
% ...

\end{sloppypar}

\begin{description}[labelsep=0.2em,align=right,labelwidth=0.7em,labelindent=0em,leftmargin=2em,noitemsep,before={\renewcommand\makelabel[1]{##1 }}]
\item[$^{1}$] Centro At\'omico Bariloche and Instituto Balseiro (CNEA-UNCuyo-CONICET), San Carlos de Bariloche, Argentina
\item[$^{2}$] Departamento de F\'\i{}sica and Departamento de Ciencias de la Atm\'osfera y los Oc\'eanos, FCEyN, Universidad de Buenos Aires and CONICET, Buenos Aires, Argentina
\item[$^{3}$] IFLP, Universidad Nacional de La Plata and CONICET, La Plata, Argentina
\item[$^{4}$] Instituto de Astronom\'\i{}a y F\'\i{}sica del Espacio (IAFE, CONICET-UBA), Buenos Aires, Argentina
\item[$^{5}$] Instituto de F\'\i{}sica de Rosario (IFIR) -- CONICET/U.N.R.\ and Facultad de Ciencias Bioqu\'\i{}micas y Farmac\'euticas U.N.R., Rosario, Argentina
\item[$^{6}$] Instituto de Tecnolog\'\i{}as en Detecci\'on y Astropart\'\i{}culas (CNEA, CONICET, UNSAM), and Universidad Tecnol\'ogica Nacional -- Facultad Regional Mendoza (CONICET/CNEA), Mendoza, Argentina
\item[$^{7}$] Instituto de Tecnolog\'\i{}as en Detecci\'on y Astropart\'\i{}culas (CNEA, CONICET, UNSAM), Buenos Aires, Argentina
\item[$^{8}$] International Center of Advanced Studies and Instituto de Ciencias F\'\i{}sicas, ECyT-UNSAM and CONICET, Campus Miguelete -- San Mart\'\i{}n, Buenos Aires, Argentina
\item[$^{9}$] Laboratorio Atm\'osfera -- Departamento de Investigaciones en L\'aseres y sus Aplicaciones -- UNIDEF (CITEDEF-CONICET), Argentina
\item[$^{10}$] Observatorio Pierre Auger, Malarg\"ue, Argentina
\item[$^{11}$] Observatorio Pierre Auger and Comisi\'on Nacional de Energ\'\i{}a At\'omica, Malarg\"ue, Argentina
\item[$^{12}$] Universidad Tecnol\'ogica Nacional -- Facultad Regional Buenos Aires, Buenos Aires, Argentina
\item[$^{13}$] University of Adelaide, Adelaide, S.A., Australia
\item[$^{14}$] Universit\'e Libre de Bruxelles (ULB), Brussels, Belgium
\item[$^{15}$] Vrije Universiteit Brussels, Brussels, Belgium
\item[$^{16}$] Centro Brasileiro de Pesquisas Fisicas, Rio de Janeiro, RJ, Brazil
\item[$^{17}$] Centro Federal de Educa\c{c}\~ao Tecnol\'ogica Celso Suckow da Fonseca, Petropolis, Brazil
\item[$^{18}$] Instituto Federal de Educa\c{c}\~ao, Ci\^encia e Tecnologia do Rio de Janeiro (IFRJ), Brazil
\item[$^{19}$] Universidade de S\~ao Paulo, Escola de Engenharia de Lorena, Lorena, SP, Brazil
\item[$^{20}$] Universidade de S\~ao Paulo, Instituto de F\'\i{}sica de S\~ao Carlos, S\~ao Carlos, SP, Brazil
\item[$^{21}$] Universidade de S\~ao Paulo, Instituto de F\'\i{}sica, S\~ao Paulo, SP, Brazil
\item[$^{22}$] Universidade Estadual de Campinas (UNICAMP), IFGW, Campinas, SP, Brazil
\item[$^{23}$] Universidade Estadual de Feira de Santana, Feira de Santana, Brazil
\item[$^{24}$] Universidade Federal de Campina Grande, Centro de Ciencias e Tecnologia, Campina Grande, Brazil
\item[$^{25}$] Universidade Federal do ABC, Santo Andr\'e, SP, Brazil
\item[$^{26}$] Universidade Federal do Paran\'a, Setor Palotina, Palotina, Brazil
\item[$^{27}$] Universidade Federal do Rio de Janeiro, Instituto de F\'\i{}sica, Rio de Janeiro, RJ, Brazil
\item[$^{28}$] Universidad de Medell\'\i{}n, Medell\'\i{}n, Colombia
\item[$^{29}$] Universidad Industrial de Santander, Bucaramanga, Colombia
\item[$^{30}$] Charles University, Faculty of Mathematics and Physics, Institute of Particle and Nuclear Physics, Prague, Czech Republic
\item[$^{31}$] Institute of Physics of the Czech Academy of Sciences, Prague, Czech Republic
\item[$^{32}$] Palacky University, Olomouc, Czech Republic
\item[$^{33}$] CNRS/IN2P3, IJCLab, Universit\'e Paris-Saclay, Orsay, France
\item[$^{34}$] Laboratoire de Physique Nucl\'eaire et de Hautes Energies (LPNHE), Sorbonne Universit\'e, Universit\'e de Paris, CNRS-IN2P3, Paris, France
\item[$^{35}$] Univ.\ Grenoble Alpes, CNRS, Grenoble Institute of Engineering Univ.\ Grenoble Alpes, LPSC-IN2P3, 38000 Grenoble, France
\item[$^{36}$] Universit\'e Paris-Saclay, CNRS/IN2P3, IJCLab, Orsay, France
\item[$^{37}$] Bergische Universit\"at Wuppertal, Department of Physics, Wuppertal, Germany
\item[$^{38}$] Karlsruhe Institute of Technology (KIT), Institute for Experimental Particle Physics, Karlsruhe, Germany
\item[$^{39}$] Karlsruhe Institute of Technology (KIT), Institut f\"ur Prozessdatenverarbeitung und Elektronik, Karlsruhe, Germany
\item[$^{40}$] Karlsruhe Institute of Technology (KIT), Institute for Astroparticle Physics, Karlsruhe, Germany
\item[$^{41}$] RWTH Aachen University, III.\ Physikalisches Institut A, Aachen, Germany
\item[$^{42}$] Universit\"at Hamburg, II.\ Institut f\"ur Theoretische Physik, Hamburg, Germany
\item[$^{43}$] Universit\"at Siegen, Department Physik -- Experimentelle Teilchenphysik, Siegen, Germany
\item[$^{44}$] Gran Sasso Science Institute, L'Aquila, Italy
\item[$^{45}$] INFN Laboratori Nazionali del Gran Sasso, Assergi (L'Aquila), Italy
\item[$^{46}$] INFN, Sezione di Catania, Catania, Italy
\item[$^{47}$] INFN, Sezione di Lecce, Lecce, Italy
\item[$^{48}$] INFN, Sezione di Milano, Milano, Italy
\item[$^{49}$] INFN, Sezione di Napoli, Napoli, Italy
\item[$^{50}$] INFN, Sezione di Roma ``Tor Vergata'', Roma, Italy
\item[$^{51}$] INFN, Sezione di Torino, Torino, Italy
\item[$^{52}$] Istituto di Astrofisica Spaziale e Fisica Cosmica di Palermo (INAF), Palermo, Italy
\item[$^{53}$] Osservatorio Astrofisico di Torino (INAF), Torino, Italy
\item[$^{54}$] Politecnico di Milano, Dipartimento di Scienze e Tecnologie Aerospaziali , Milano, Italy
\item[$^{55}$] Universit\`a del Salento, Dipartimento di Matematica e Fisica ``E.\ De Giorgi'', Lecce, Italy
\item[$^{56}$] Universit\`a dell'Aquila, Dipartimento di Scienze Fisiche e Chimiche, L'Aquila, Italy
\item[$^{57}$] Universit\`a di Catania, Dipartimento di Fisica e Astronomia ``Ettore Majorana``, Catania, Italy
\item[$^{58}$] Universit\`a di Milano, Dipartimento di Fisica, Milano, Italy
\item[$^{59}$] Universit\`a di Napoli ``Federico II'', Dipartimento di Fisica ``Ettore Pancini'', Napoli, Italy
\item[$^{60}$] Universit\`a di Palermo, Dipartimento di Fisica e Chimica ''E.\ Segr\`e'', Palermo, Italy
\item[$^{61}$] Universit\`a di Roma ``Tor Vergata'', Dipartimento di Fisica, Roma, Italy
\item[$^{62}$] Universit\`a Torino, Dipartimento di Fisica, Torino, Italy
\item[$^{63}$] Benem\'erita Universidad Aut\'onoma de Puebla, Puebla, M\'exico
\item[$^{64}$] Unidad Profesional Interdisciplinaria en Ingenier\'\i{}a y Tecnolog\'\i{}as Avanzadas del Instituto Polit\'ecnico Nacional (UPIITA-IPN), M\'exico, D.F., M\'exico
\item[$^{65}$] Universidad Aut\'onoma de Chiapas, Tuxtla Guti\'errez, Chiapas, M\'exico
\item[$^{66}$] Universidad Michoacana de San Nicol\'as de Hidalgo, Morelia, Michoac\'an, M\'exico
\item[$^{67}$] Universidad Nacional Aut\'onoma de M\'exico, M\'exico, D.F., M\'exico
\item[$^{68}$] Institute of Nuclear Physics PAN, Krakow, Poland
\item[$^{69}$] University of \L{}\'od\'z, Faculty of High-Energy Astrophysics,\L{}\'od\'z, Poland
\item[$^{70}$] Laborat\'orio de Instrumenta\c{c}\~ao e F\'\i{}sica Experimental de Part\'\i{}culas -- LIP and Instituto Superior T\'ecnico -- IST, Universidade de Lisboa -- UL, Lisboa, Portugal
\item[$^{71}$] ``Horia Hulubei'' National Institute for Physics and Nuclear Engineering, Bucharest-Magurele, Romania
\item[$^{72}$] Institute of Space Science, Bucharest-Magurele, Romania
\item[$^{73}$] Center for Astrophysics and Cosmology (CAC), University of Nova Gorica, Nova Gorica, Slovenia
\item[$^{74}$] Experimental Particle Physics Department, J.\ Stefan Institute, Ljubljana, Slovenia
\item[$^{75}$] Universidad de Granada and C.A.F.P.E., Granada, Spain
\item[$^{76}$] Instituto Galego de F\'\i{}sica de Altas Enerx\'\i{}as (IGFAE), Universidade de Santiago de Compostela, Santiago de Compostela, Spain
\item[$^{77}$] IMAPP, Radboud University Nijmegen, Nijmegen, The Netherlands
\item[$^{78}$] Nationaal Instituut voor Kernfysica en Hoge Energie Fysica (NIKHEF), Science Park, Amsterdam, The Netherlands
\item[$^{79}$] Stichting Astronomisch Onderzoek in Nederland (ASTRON), Dwingeloo, The Netherlands
\item[$^{80}$] Universiteit van Amsterdam, Faculty of Science, Amsterdam, The Netherlands
\item[$^{81}$] Case Western Reserve University, Cleveland, OH, USA
\item[$^{82}$] Colorado School of Mines, Golden, CO, USA
\item[$^{83}$] Department of Physics and Astronomy, Lehman College, City University of New York, Bronx, NY, USA
\item[$^{84}$] Michigan Technological University, Houghton, MI, USA
\item[$^{85}$] New York University, New York, NY, USA
\item[$^{86}$] University of Chicago, Enrico Fermi Institute, Chicago, IL, USA
\item[$^{87}$] University of Delaware, Department of Physics and Astronomy, Bartol Research Institute, Newark, DE, USA
\item[] -----
\item[$^{a}$] Max-Planck-Institut f\"ur Radioastronomie, Bonn, Germany
\item[$^{b}$] also at Kapteyn Institute, University of Groningen, Groningen, The Netherlands
\item[$^{c}$] School of Physics and Astronomy, University of Leeds, Leeds, United Kingdom
\item[$^{d}$] Fermi National Accelerator Laboratory, Fermilab, Batavia, IL, USA
\item[$^{e}$] Pennsylvania State University, University Park, PA, USA
\item[$^{f}$] Colorado State University, Fort Collins, CO, USA
\item[$^{g}$] Louisiana State University, Baton Rouge, LA, USA
\item[$^{h}$] now at Graduate School of Science, Osaka Metropolitan University, Osaka, Japan
\item[$^{i}$] Institut universitaire de France (IUF), France
\item[$^{j}$] now at Technische Universit\"at Dortmund and Ruhr-Universit\"at Bochum, Dortmund and Bochum, Germany
\end{description}

\section*{Acknowledgments}

\begin{sloppypar}
The successful installation, commissioning, and operation of the Pierre
Auger Observatory would not have been possible without the strong
commitment and effort from the technical and administrative staff in
Malarg\"ue. We are very grateful to the following agencies and
organizations for financial support:
\end{sloppypar}

\begin{sloppypar}
Argentina -- Comisi\'on Nacional de Energ\'\i{}a At\'omica; Agencia Nacional de
Promoci\'on Cient\'\i{}fica y Tecnol\'ogica (ANPCyT); Consejo Nacional de
Investigaciones Cient\'\i{}ficas y T\'ecnicas (CONICET); Gobierno de la
Provincia de Mendoza; Municipalidad de Malarg\"ue; NDM Holdings and Valle
Las Le\~nas; in gratitude for their continuing cooperation over land
access; Australia -- the Australian Research Council; Belgium -- Fonds
de la Recherche Scientifique (FNRS); Research Foundation Flanders (FWO),
Marie Curie Action of the European Union Grant No.~101107047; Brazil --
Conselho Nacional de Desenvolvimento Cient\'\i{}fico e Tecnol\'ogico (CNPq);
Financiadora de Estudos e Projetos (FINEP); Funda\c{c}\~ao de Amparo \`a
Pesquisa do Estado de Rio de Janeiro (FAPERJ); S\~ao Paulo Research
Foundation (FAPESP) Grants No.~2019/10151-2, No.~2010/07359-6 and
No.~1999/05404-3; Minist\'erio da Ci\^encia, Tecnologia, Inova\c{c}\~oes e
Comunica\c{c}\~oes (MCTIC); Czech Republic -- GACR 24-13049S, CAS LQ100102401,
MEYS LM2023032, CZ.02.1.01/0.0/0.0/16{\textunderscore}013/0001402,
CZ.02.1.01/0.0/0.0/18{\textunderscore}046/0016010 and
CZ.02.1.01/0.0/0.0/17{\textunderscore}049/0008422 and CZ.02.01.01/00/22{\textunderscore}008/0004632;
France -- Centre de Calcul IN2P3/CNRS; Centre National de la Recherche
Scientifique (CNRS); Conseil R\'egional Ile-de-France; D\'epartement
Physique Nucl\'eaire et Corpusculaire (PNC-IN2P3/CNRS); D\'epartement
Sciences de l'Univers (SDU-INSU/CNRS); Institut Lagrange de Paris (ILP)
Grant No.~LABEX ANR-10-LABX-63 within the Investissements d'Avenir
Programme Grant No.~ANR-11-IDEX-0004-02; Germany -- Bundesministerium
f\"ur Bildung und Forschung (BMBF); Deutsche Forschungsgemeinschaft (DFG);
Finanzministerium Baden-W\"urttemberg; Helmholtz Alliance for
Astroparticle Physics (HAP); Helmholtz-Gemeinschaft Deutscher
Forschungszentren (HGF); Ministerium f\"ur Kultur und Wissenschaft des
Landes Nordrhein-Westfalen; Ministerium f\"ur Wissenschaft, Forschung und
Kunst des Landes Baden-W\"urttemberg; Italy -- Istituto Nazionale di
Fisica Nucleare (INFN); Istituto Nazionale di Astrofisica (INAF);
Ministero dell'Universit\`a e della Ricerca (MUR); CETEMPS Center of
Excellence; Ministero degli Affari Esteri (MAE), ICSC Centro Nazionale
di Ricerca in High Performance Computing, Big Data and Quantum
Computing, funded by European Union NextGenerationEU, reference code
CN{\textunderscore}00000013; M\'exico -- Consejo Nacional de Ciencia y Tecnolog\'\i{}a
(CONACYT) No.~167733; Universidad Nacional Aut\'onoma de M\'exico (UNAM);
PAPIIT DGAPA-UNAM; The Netherlands -- Ministry of Education, Culture and
Science; Netherlands Organisation for Scientific Research (NWO); Dutch
national e-infrastructure with the support of SURF Cooperative; Poland
-- Ministry of Education and Science, grants No.~DIR/WK/2018/11 and
2022/WK/12; National Science Centre, grants No.~2016/22/M/ST9/00198,
2016/23/B/ST9/01635, 2020/39/B/ST9/01398, and 2022/45/B/ST9/02163;
Portugal -- Portuguese national funds and FEDER funds within Programa
Operacional Factores de Competitividade through Funda\c{c}\~ao para a Ci\^encia
e a Tecnologia (COMPETE); Romania -- Ministry of Research, Innovation
and Digitization, CNCS-UEFISCDI, contract no.~30N/2023 under Romanian
National Core Program LAPLAS VII, grant no.~PN 23 21 01 02 and project
number PN-III-P1-1.1-TE-2021-0924/TE57/2022, within PNCDI III; Slovenia
-- Slovenian Research Agency, grants P1-0031, P1-0385, I0-0033, N1-0111;
Spain -- Ministerio de Ciencia e Innovaci\'on/Agencia Estatal de
Investigaci\'on (PID2019-105544GB-I00, PID2022-140510NB-I00 and
RYC2019-027017-I), Xunta de Galicia (CIGUS Network of Research Centers,
Consolidaci\'on 2021 GRC GI-2033, ED431C-2021/22 and ED431F-2022/15),
Junta de Andaluc\'\i{}a (SOMM17/6104/UGR and P18-FR-4314), and the European
Union (Marie Sklodowska-Curie 101065027 and ERDF); USA -- Department of
Energy, Contracts No.~DE-AC02-07CH11359, No.~DE-FR02-04ER41300,
No.~DE-FG02-99ER41107 and No.~DE-SC0011689; National Science Foundation,
Grant No.~0450696, and NSF-2013199; The Grainger Foundation; Marie
Curie-IRSES/EPLANET; European Particle Physics Latin American Network;
and UNESCO.
\end{sloppypar}


\end{document}